\documentclass[12pt]{iopart}
\usepackage{bm}
\usepackage{graphicx}
\usepackage{cite}
\usepackage{color}
\usepackage[binary,amssymb]{SIunits}

\usepackage{color} 
\usepackage{cancel}
\begin{document}
\title{Twisting Dirac fermions: Circular dichroism in bilayer graphene}

\author{$^1$E. Su\'{a}rez Morell, $^2$Leonor Chico, $^2$Luis Brey}
\address{$^1$Departamento de F\'{i}sica, Universidad T\'{e}cnica
Federico Santa Mar\'{i}a, Casilla 110-V, Valpara\'{i}so, Chile}
\address{$^2$Departamento de Teor\'{\i}a y Simulaci\'on de Materiales, Instituto de Ciencia de Materiales de Madrid, Consejo Superior de Investigaciones Cient\' ificas ICMM-CSIC, 28049 Madrid, Spain}

\date{\today}

\begin{abstract}
Twisted bilayer graphene is a chiral system which has been recently shown to present circular dichroism. In this work we show that the origin of this optical activity is the rotation of the Dirac fermions' helicities in the top and bottom layer. 
Starting from the Kubo formula, we obtain a compact expression for the Hall conductivity that takes into account the dephasing of the electromagnetic field between the top and bottom layers and gathers all the symmetries of the system. Our results are based in both a continuum and a tight-binding model, and they can be generalized to any two-dimensional Dirac material with a chiral stacking between layers. 

\end{abstract}

\maketitle

\section{Introduction}
Circular dichroism (CD) is the difference in the optical response of materials to left- and right-handed circularly polarized light 
\cite{Rodger_book,Barron_book}.
  CD is intimately related to chirality; a system is chiral if it is distinguishable from its mirror image. 
  The study of chiral molecules is an important branch of stereochemistry \cite{Inoue_book}; 
  in principle, chiral materials may be used to realize circular polarizers \cite{Barron_book} as well as for applications in spintronics \cite{Okamoto:2014aa} and valleytronics \cite{Yao:2008aa,Aivazian:2015aa, Srivastava:2015aa}. The integration of circular polarizers in nanodevices requires the realization of planar and atomically thin crystal structures, but these systems in general show weaker circular dichroism because an atomically thin film is inherently achiral. 
  However, the design of metamaterials with coupled planar components in a chiral three-dimensional arrangement has resulted in large CD 
  \cite{Kuwata-Gonokami:2005aa,Rogacheva:2006aa,Decker:2007aa,Zhao:2012aa}. 
 Additionally, the concept of extrinsic chirality is being applied to 
achiral planar 
 artificial nanostructures
 that produce a large circular dichroism if the light incides in certain directions 
  \cite{Plum:2009}. 
 However, such metamaterials do not provide insights of how to make atomically thick polarizers with intrinsic chirality. 

Recently, a strong circular dichroism has been reported in two-atom-thick twisted bilayer graphene (TBG) \cite{Kim:2016aa}. 
It has been experimentally observed that, by changing the magnitude and sign of the interlayer rotation, it is possible to tune the energy, CD intensity and polarity.
TBG consists of two graphene layers stacked with a rotational mismatch \cite{Castro_2007}, characterized by a relative rotation angle $\theta$ with respect to a perfect AB stacking, in which only half of all the carbon atoms on opposite graphene sheets are vertically linked \cite{Guinea_2009,Katsnelson-book}.
TBG can be obtained by different preparation techniques, such as growth on a SiC substrate, chemical vapor deposition or folding a single layer graphene \cite{Rozhkov:2016aa}.  Exfoliation can also yield bilayer graphene rotated an arbitrary angle $\theta$.

\begin{figure}[htbp]
\centering 
\includegraphics[width=9cm,clip]{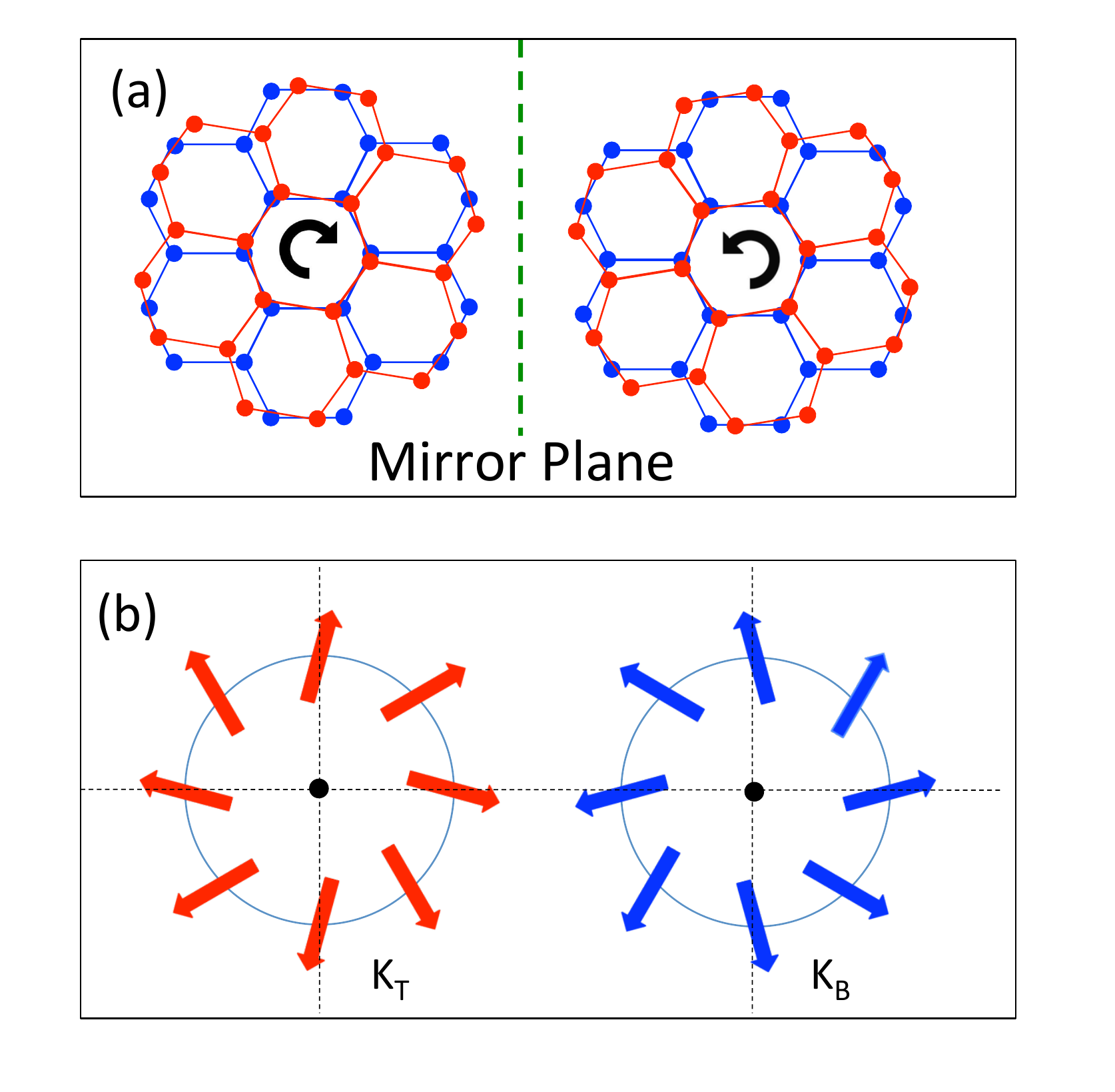}
\caption{(Color online) (a) Portion of the atomic structure of two graphene layers rotated  clockwise (left) and 
anticlockwise (right).
Blue (red) circles correspond to atoms in the bottom (top) layer.  
Clockwise- and anticlockwise-rotated bilayer graphene are distinguishable; they are 
 mirror images.  Therefore, twisted bilayer graphene 
is a chiral material. (b) Expectation values of the sublattice pseudospin near the Dirac points of the T and B graphene layers. The pseudospin in the T layer 
is rotated an angle $\theta$ with respect to the pseudospin in the B layer.
 } 

\label{figure1}
\end{figure}

Each graphene monolayer is optically isotropic; therefore, 
CD in TBG cannot be explained as in the 
case of chirality in metamaterials, in which the two constituent layers have orthogonal electric dipoles that are elastically coupled \cite{Yin:2013aa,Rogacheva:2006aa,Zheludev_book}.
In absence of magnetic impurities or external fields, the occurrence of intrinsic CD is connected to the chirality of the system, i.e., 
to the fact that a system cannot be superposed onto its mirror image. In the case of TBG, 
the combination of rotation and finite separation between the layers makes the mirror image of a TBG with twist angle  $\theta$  equal to 
 TBG rotated an angle $-\theta$, see Fig. \ref{figure1}(a).

As it is well-known, the electronic properties of monolayer graphene near the Fermi energy are governed by the massless Dirac equation,
$H_0=v_F {\bf k } \cdot {\bm \tau}$, 
where $v_F$ is the Fermi velocity, $\bf k$ is the two-dimensional wavevector and the matrices {\boldmath{$\tau$}} are Pauli matrices acting on the $A$ and $B$ sublattice 
basis. 
Dirac fermions are chiral in
the sense 
 that the expectation value of the sublattice pseudospin is parallel or antiparallel to the momentum operator, i.e., they have a well-defined helicity.
For a TBG with a rotation angle $\theta$, the Dirac cones of the top (T) and bottom (B) graphene layers 
are separated by a wavevector $K_{\theta}$=$k_D \sin \theta$, where $k_D$=$4 \pi/ 3a$ and $a$ is the
graphene lattice parameter \cite{Lopes_2007,Mele_2010,Lopes_2012,Bistritzer2011,SanJose_2012}. 
Additionally, electrons can tunnel 
between the T and B layers; this produces a renormalization of the Fermi velocity at the Dirac points \cite{Song_2010,Luican_2011} and the occurrence of van Hove singularities close to the Fermi energy at low rotation angles \cite{Li:2010aa,Trambly_2012b,Havener:2014aa}. 

The twist between the layers also 
implies 
a rotation 
of the Pauli matrices  {\boldmath{$\tau$}} in the Dirac Hamiltonian of one layer with respect to the other.
Therefore, the chirality (helicity) in one of the layers is rotated an angle $\theta$ with respect to the chirality in the other; see Fig. \ref{figure1}(b).
Such rotation 
only has a parametrically small effect on the energy position of the van Hove singularity and on the Fermi velocity renormalization \cite{Bistritzer2011}. 
However, we find that {\it  the rotation of the Dirac fermions' chirality in a layer with respect to the other is the 
 actual 
responsible for the occurrence of CD in TBG.} This is one of the main conclusions of the present work. 
In the following we 
vindicate 
this conclusion and present results for the  optical conductivity  and  circular dichroism in TBG, based in the Kubo formalism within the continuous Dirac-like and tight-binding Hamiltonians.

\begin{figure}[htbp]
\centering
\includegraphics[width=9cm,clip]{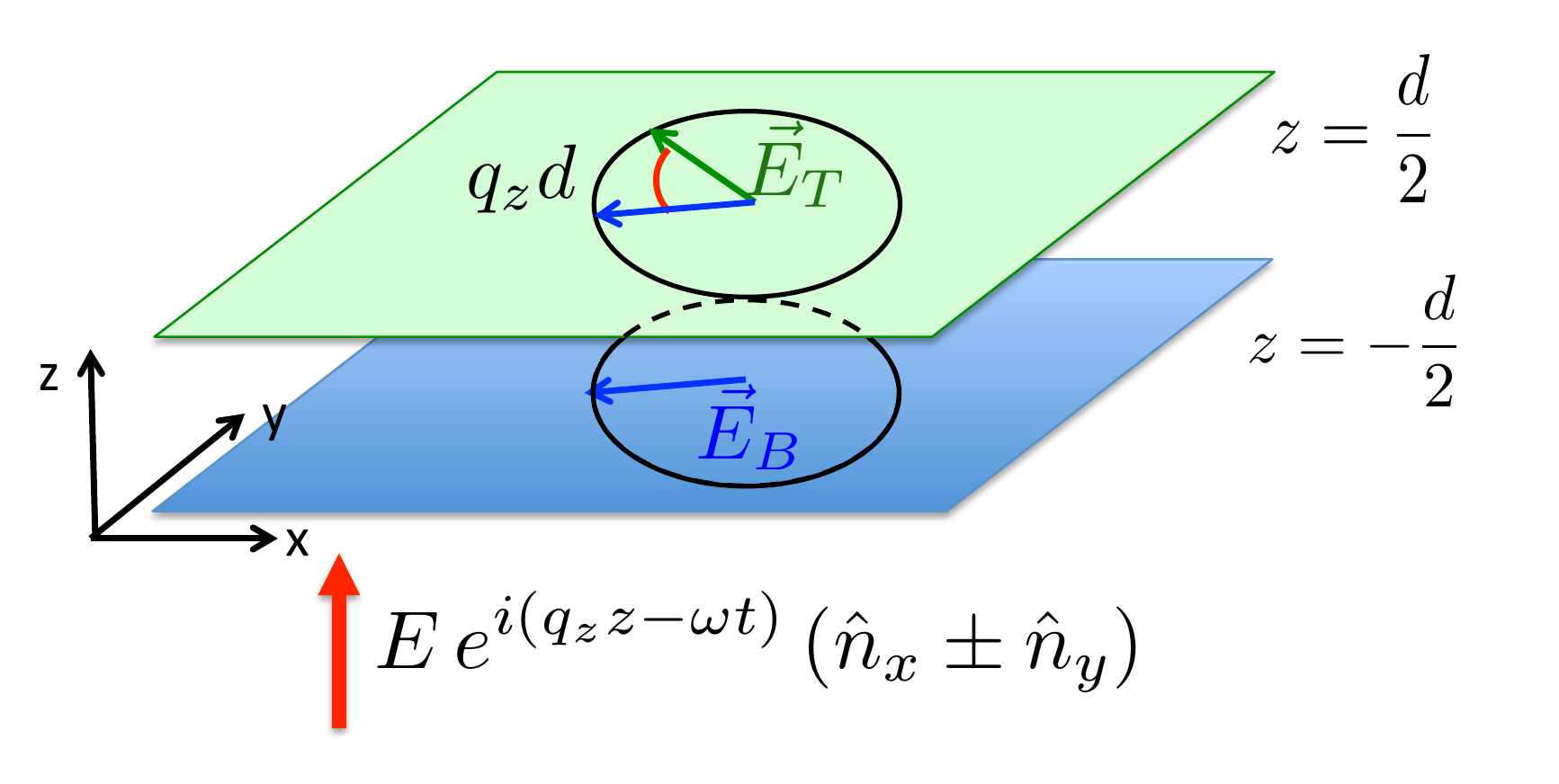}
\caption{(Color online) A  circularly polarized light with momentum $q_z$ and frequency $\omega$ moving in the  $z$-direction
 is incident on  
 a TBG. 
The separation $d$ between the two graphene sheets produces a dephasing $q_zd$ between the electric field of the light on the top and bottom layers. } 
\label{figure2}
\end{figure}

\section{Ellipticity and Hall conductivity} 
The change in the polarization of the light  when  it is transmitted through a thin film is characterized by the ellipticity $\psi$, 
defined as $\psi=(I_L-I_R)/2(I_L+I_R)$, where $I_L$ and $I_R$ are the absorption of left and right circularly polarized light, respectively \cite{Barron_book}. 
The optical absorption is obtained from the reflection and transmission matrices, which are derived from the application of the boundary conditions for the electromagnetic field  \cite{Oliva-Leyva:2015aa}. 
Considering the thin film as a two-dimensional system separating two media, the ellipticity takes the form 
\begin{equation}
\psi= 
\frac {{\rm Im} (\sigma _{x,y}) } {2\,{\rm Re} (\sigma_{x,x})}, 
\end{equation}
 where $\sigma _{x,x}$ and $\sigma _{x,y}$
are the longitudinal and Hall conductivities  of the two-dimensional system, respectively.

Strictly planar systems do not show intrinsic circular dichroism; in order to correctly describe 
 the optical absorption in TBG, it is essential to 
take into account the separation between the two graphene sheets $d=3.35$\AA, and the dephasing of the electric field of the light on the top and bottom layers; see Fig. \ref{figure2}.  The  current in the $\alpha$=$x,y$ direction in the TBG system has two components,
\begin{equation}
J_\alpha= J^T_\alpha e^{iq_z\frac{d}{2} }+  J^B_\alpha e^{-iq_z\frac{d}{2}} 
 \, \, \, ,
\label{Total-Current}
\end{equation}
where ${\bf J}^{T(B)}$ is the current created in the T (B) layer by the electric field of the incident light evaluated at the center of the bilayer.
We compute the optical conductivity using the Kubo formula \cite{Allen_book},
\begin{equation}
\sigma_{\mu,\nu}(\omega) = i g _s \frac{\hbar}{S} \sum_{ ij} \frac{f(\varepsilon_i) - f(\varepsilon_j)}{\varepsilon_j - \varepsilon_i} 
\frac{\langle i | J_\mu | j \rangle  \langle j | J_\nu | i \rangle }{ \hbar( \omega + i \eta) +  {\varepsilon_i - \varepsilon_j} } \, \, \, ,
\label{Kubo}
 \end{equation}
 here $S$ is the sample area, $g_s$=2 is the spin degeneracy, $f(\epsilon)$ is the Fermi function, $|i>$ are eigenstates of the total Hamiltonian with eigenenergies $\varepsilon _i$, 
  and $\eta$ is a phenomenological lifetime 
  that takes into account the effect of lattice vibrations, i.e., electron-phonon coupling, and the lifetime of the states involved in the transition 
   \cite{Alivisatos:1988}. We take $\eta=0.025$ eV for the calculations of the optical conductivity.   
 Expanding this expression up to linear order in the phase shift $q_z d$, 
\begin{eqnarray}
\sigma_{\mu,\mu} (\omega,q_z)& = &  \sigma_{\mu,\mu} ^{(0)} (\omega) + \cancelto{0}{ \sigma_{\mu,\mu} ^{(1)}(\omega,q_z)} + ...\\
\nonumber
\sigma_{x,y} (\omega,q_z)& =&   \cancelto{0}{ \sigma_{x,y} ^{(0)}} +  \sigma_{x,y} ^{(1)} (\omega,q_z) +   ...
\label{sigmas}
\end{eqnarray}
here $\sigma_{\mu,\mu} ^{0}(\omega) $ is the diagonal optical conductivity of the TBG \cite{Moon:13,Tabert:13,Stauber:2013aa}. The diagonal part of the conductivity has no terms linear in $q_zd$, indicating the absence of linear dichroism in the system and the preservation of time reversal symmetry. 
At order zero in $q_zd$, $\sigma_{x,y}$ vanishes  because in the absence of magnetism, a planar atomic monolayer does not 
present circular dichroism. Finally, the lowest nonzero order of the Hall conductivity takes the form 
\begin{equation}
\sigma_{x,y}^{(1)}(\omega,q_z) = - 2 g _s {q_z d}  \frac{\hbar}{S}  \sum_{ i,j}  \frac{f(\varepsilon_i)-f(\varepsilon_j)}{\varepsilon_j-\varepsilon_i} 
\frac{\langle i | J_x ^T | j \rangle  \langle j| J_y ^B | i\rangle 
}{ \hbar( \omega + i \eta) +  {\varepsilon_i - \varepsilon_j} }\, \, \,.
\label{sigmaxy}
\end{equation}
Because of time reversal symmetry, the dynamical Hall conductivity satisfies the following relations,
$\sigma _{x,y}(\omega,q_z)$=$-\sigma _{y,x} (\omega,q_z)$, $\sigma _{x,y}(\omega,q_z)$=$-\sigma _{x,y} (\omega,-q_z)$ and 
$\sigma _{x,y}(\omega=0,q_z)=0.$ 
Eq. (\ref{sigmaxy}) is a Hall drag-like 
conductivity, 
indicating the current in one of the layers induced by a current flowing in the perpendicular direction 
on the opposite layer. 

\section{Dirac-like continuous Hamiltonian}
In order to evaluate the Hall conductivity, we need the eigenvalues and eigenvectors of the TBG.  
Although TBG systems show moir\'{e} patterns that look periodic \cite{Trambly_2012b}, most of them are incommensurate structures. Only for a discrete set of twist angles the systems are periodic, 
being possible to perform atomistic calculations for the obtention of  
their band structures \cite{Shallcross:2010aa,Morell_2010,Trambly-de-Laissardiere:2010aa,Morell_2011b,Sato_2012,Landgraf:2013aa,Suarez-Morell:2014aa}. 
Several semi-analytical theories based in the continuum Dirac model have been developed to describe the low energy physics of TBG with small rotation angles \cite{Lopes_2007,Mele_2010,Lopes_2012,Bistritzer2011,SanJose_2012}. 
The ingredients of the continuum model are the following: one Dirac Hamiltonian for each graphene layer, $H_T$ and $H_B$, and an interlayer tunneling term that contains the two Fourier components of the tunneling with moments 
${\bf G} _1$=$\frac 1 2 (\sqrt{3},3)K_{\theta}$ and
${\bf G}_2$=$\frac 1 2 (-\sqrt{3},3)K_{\theta}$, where the coupling is larger, 
\begin{equation}
 H (\theta, {\bf k})\!= \! \left (\!  \begin{array}{cc}
H_T & \bar  T_0+ \sum _{i=1,2} {\bar T _i e ^{i {\vec G}_i \vec{r}}} \\
\bar T_0+ \sum _{i=1,2} \bar T _i ^{\dagger} e ^{-i {\vec G}_i \vec{r}} & H_B
\end{array}
\! \right  ) ,
\label{htotal}
\end{equation}
the tunneling matrices have the form 
 $\bar T_0$=$\bar \omega
\left ( \begin{array}{cc} 1 & 1 \\ 1 & 1\end{array} \right )$, $\bar T_1=  z^* e ^{i \frac {\pi} 3 \tau_z} \bar T_0 e ^{-i \frac {\pi} 3 \tau_z}$, and 
$\bar  T_2$=${\bar T_1} ^*$, where $z=e^{i \frac {2\pi} 3}$ and $\bar \omega$ is the hopping amplitude  \cite{Lu:2014aa}. 
The tunneling term describes the spatial modulation of the interlayer hopping produced by the misalignment of the carbon atoms in the T and B layers.
The  modulation of the hopping makes the system periodic  and  the electronic structure shows Bloch bands  in a Brillouin zone defined by the  moments
${\bf G} _1$ and ${\bf G} _2$. In real space, the unit cell corresponds to the moir\'{e} pattern of the TBG.  The diagonal terms, $H_T$ and $H_B$  in Eq. (\ref{htotal}), are Dirac  Hamiltonians centered  at the displaced Dirac points $\pm {\bf K}_{\theta} /2$, with ${\bf K}_{\theta} $=$(0, K_{\theta})$,
\begin{equation}
H_{T,B} = \hbar v_F \, {\bm \tau} ^{\mp\theta /2} \cdot \left ({\bf k} \mp \frac {{\bf K_{\theta} }}2 \right) \, \, ,
\label{HTB}
\end{equation}
the matrices $ {\bm \tau} ^{\mp\theta /2}$ are defined  as $ {\bm \tau} ^{\theta }= e ^{i \theta \, \tau_z/2} {\bm \tau}e ^{-i \theta \,  \tau_z/2}$, and appear in order to describe the rotation  of the
Dirac Hamiltonians with respect to a fixed coordinate system. We take $v_F=\frac{\sqrt{3} \gamma_0 a}{2}$  and  the permittivity $\epsilon_r=1$. For small twist angles, electronic states 
from different valleys of the same layer
 do not mix, 
 and therefore in the continuum model 
 spin and valley degeneracy holds.

\begin{figure}[htbp]
\centering
\includegraphics[width=9cm,clip]{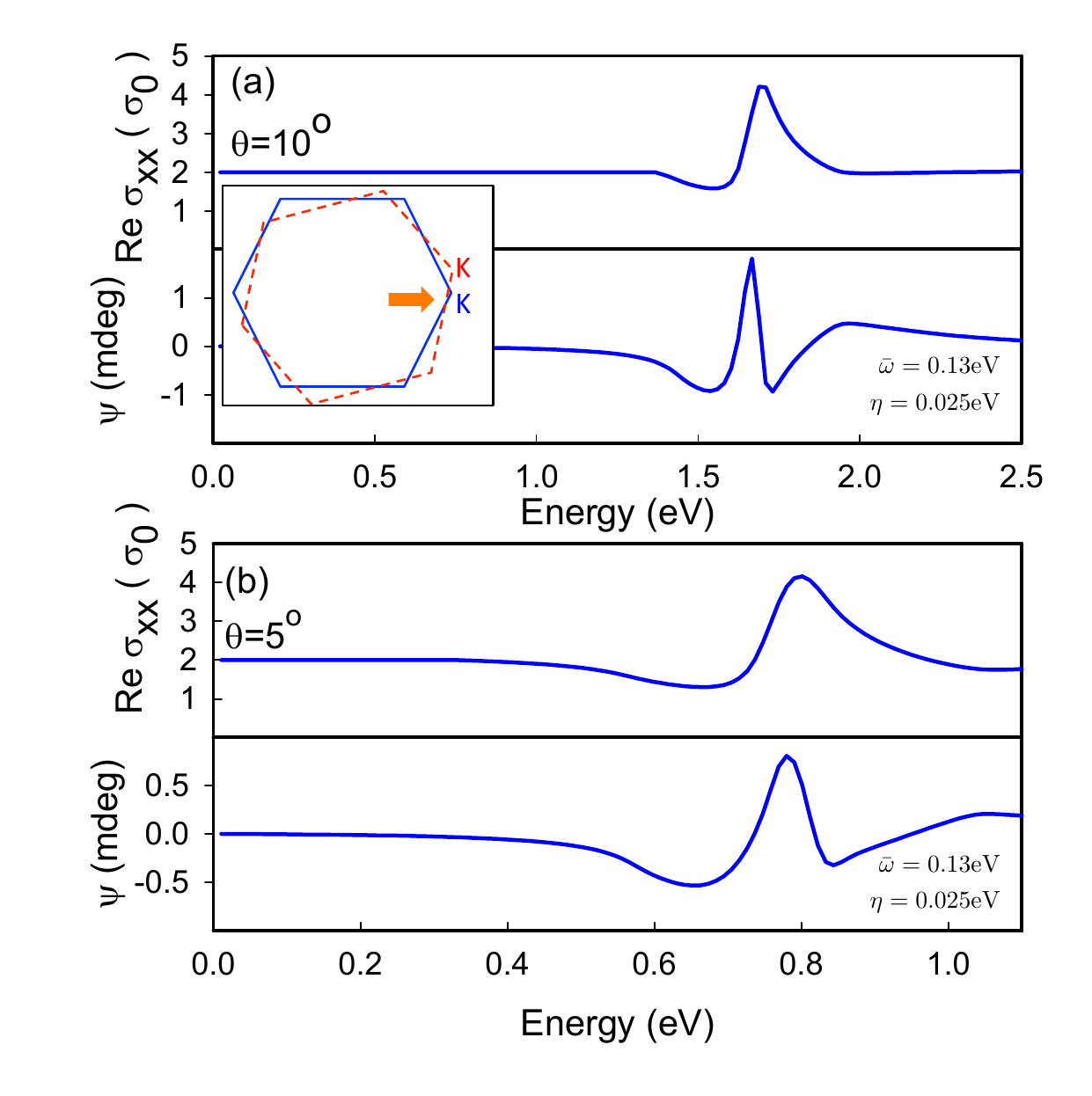}
\caption{(Color online) Optical conductivity and ellipticity for a TBG with a rotation angle (a) $\theta$=5$^ {\rm o}$ and (b) $\theta$=10$^ {\rm o}$, as obtained with the continuous Dirac-like Hamiltonian.
In the inset we show the Brillouin zones of the two rotated graphene layers. Near the Dirac points arising from opposite layers, the T and B electronic bands cross and create van Hove singularities,
which reflects in the  optical conductivity.  In the calculation we use  $\eta$=0.025 eV and  $\sigma _0$=$\frac{ 1 }{4} \frac {e^2 } {\hbar}$.} 
\label{figure3}
\end{figure}

In Fig. \ref{figure3} we plot the real part of the diagonal conductivity and the ellipticity of a TBG with rotation angles (a) $\theta$=10$^{\rm o}$  and (b) $\theta$=5$^{\rm o}$. The conductivity is plotted in units of the universal dynamical conductivity of a graphene monolayer 
$\sigma _0$=$\frac {g_s g_v}{16} \frac {e^2 } {\hbar}$ \cite{Nair06062008}, where $g_v$=2 is the valley degeneracy. Note that the ellipticity is nearly proportional to the Hall conductivity. At low frequencies the conductivity is practically constant and equal to 
 twice the optical absorption of a monolayer. At energies of the order of $\hbar v_F K_{\theta}$ the conductivity exhibits a dip-peak structure; it is a signature of the van Hove singularity that occurs 
at the intersection  near the Dirac points of the bands 
arising from different layers \cite{Tabert:13,Moon:13,Stauber:2013aa}. Likewise, the ellipticity is zero at low frequencies and shows a peak-dip 
structure at energies near 
 the van Hove singularities.   In the following we 
  analyze the origin and form of the imaginary part of the Hall conductivity in TBG.

\begin{figure}[htbp]
\centering
\includegraphics[width=9cm,clip]{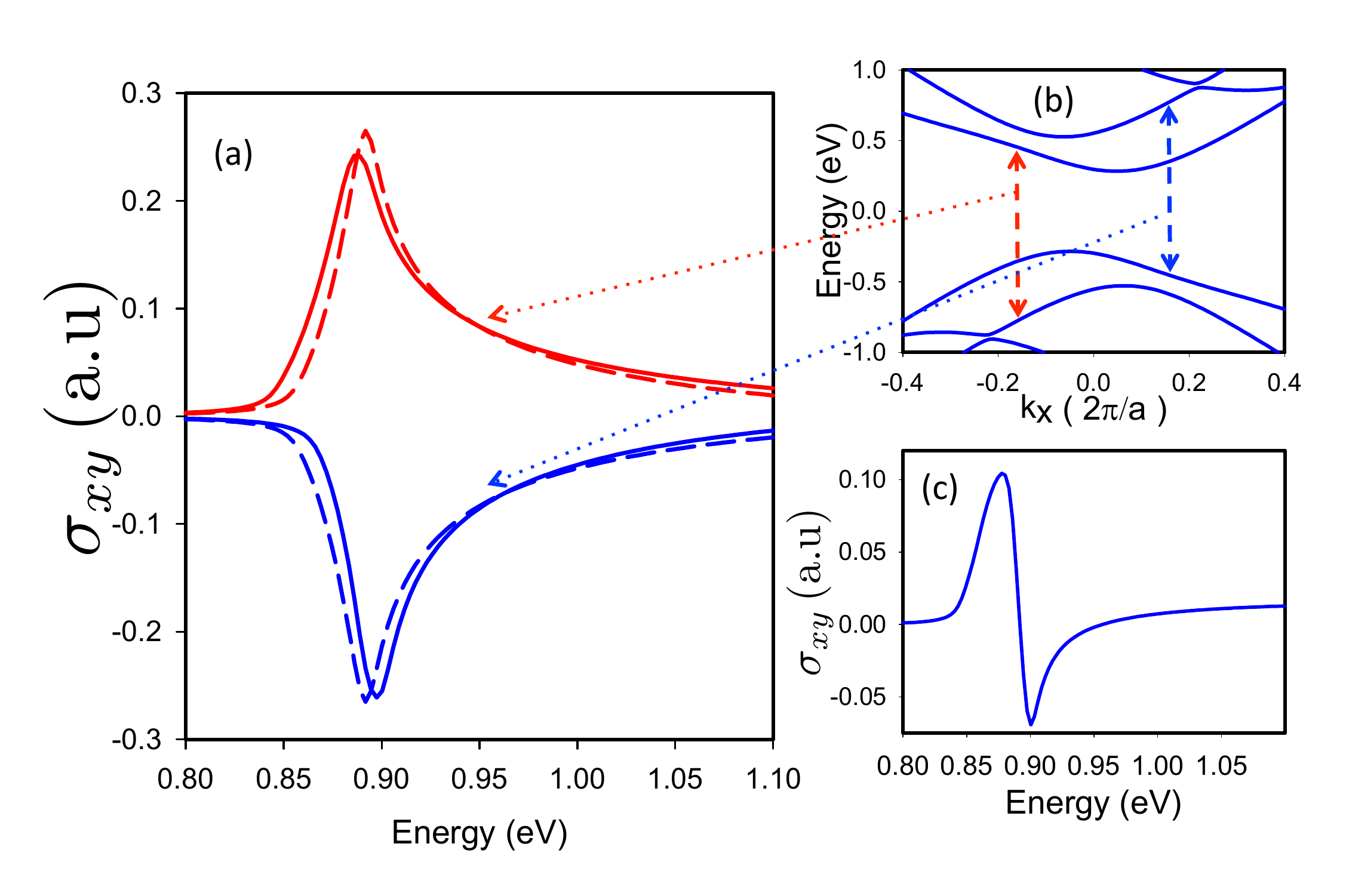}
\caption{(Color online)  (a) Contribution to the Hall conductivity due to transitions from the second highest energy valence band to the lowest energy conduction band (red),  and the opposite transitions (blue). Dashed lines correspond to the case where the rotation of the sublattice pseudospin  of the T and B  Dirac electrons is neglected.
In that case, the two contributions cancel and the Hall conductivity is zero. (b) Band structure of TBG along the $k_x$ direction with $k_y$=0. When the  rotation of the chiral symmetry is taken into account, the two intersubband contributions (continuous line in (a)) do not cancel; they are slightly shifted in energy and this results in a finite Hall conductivity.
The sum of shifted energy peaks with opposite sign yields a Hall conductivity with a peak-dip form. 
The results correspond to $\theta$=5$^{\rm o}$, $\bar \omega$=0.13 eV, and $\eta$=0.005 eV.} 
\label{Figure4}
\end{figure}

The  rotation of the pseudospin orientation in the top layer with respect to the orientation in the bottom layer, Fig. \ref{figure1}, is a consequence of the real space chiral symmetry in TBG,
see Fig. \ref{figure1}(a).  If the rotation of the Pauli matrices in the Hamiltonian of the TBG  (Eq. (\ref{htotal}))  is neglected,  then the Hamiltonian satisfies the following relations: 
\begin{eqnarray}
&& \left ( \! \begin{array}{cc} 0 & \tau_x \\ \tau _x & 0 \end{array} \! \right )  \, H _{0} (\theta,-{\bf k}) \, \left ( \!  \begin{array}{cc} 0 & \tau _x \\ \tau _x & 0 \end{array}\!  \right )   \!= \! 
H_{0} ^* (\theta,k_x,-k_y) \nonumber \\
&& \left ( \!  \begin{array}{cc} 0 &i  \tau_x \\ -i \tau _x & 0 \end{array} \! \right )  \, H _{0} (\theta,{\bf k}) \, \left ( \! \begin{array}{cc} 0 & i \tau _x \\ -i \tau _x & 0 \end{array} \! \right )  \!= \!  
-\!H_{0} ^* (\theta,-{\bf k}) 
\label{symmetries}
\end{eqnarray}
where $H_0$ is the Hamiltonian Eq. (\ref{htotal}) evaluated with   $ {\bm \tau} ^{\mp \theta/2 } \rightarrow {\bm \tau}$.
These symmetries imply that the eigenvalues of $H_0$ satisfy the relations
\begin{eqnarray}
\varepsilon _0 ({\bf k}) & = &  -\varepsilon _0 (-{\bf k}) \, \, \, \, {\rm and}  \nonumber \\
\varepsilon _0 ({\bf k}) & = &  \varepsilon _0 (k_x,-k_y)  .
\end{eqnarray}
Armed with these symmetries, it is possible to prove that the contribution to the Hall conductivity of a transition from a valence band  with energy $E_v$ to a conduction band  with energy $E_c$ is equal and of opposite sign to the contribution from a valence band with energy $-E_c$ to a conduction band with energy $-E_v$, 
\begin{equation}
\sigma _{x,y \, (E_v,E_c)} ^{(1)} (\omega,q_z) = - \sigma _{x,y \, (-E_c,-E_v)}^{(1)}  (\omega,q_z), 
\end{equation}
 therefore, when summing over all  pairs of bands the Hall conductivity vanishes, $\sigma _{x,y} ^{(1)}(\omega,q_z)=0.$ This is shown in Fig. \ref{Figure4}(a), where the contribution from 
the second highest   energy valence  band to the lowest energy conduction band, Fig. \ref{Figure4}(b), cancels exactly the contribution from the pair of bands with opposite energies. When the relative rotation of the sublattice pseudospin between graphene layers is considered, the 
contributions do not cancel (see Fig. \ref{Figure4}(a)), and this results in a finite Hall conductivity with a peak-dip form, Fig. \ref{Figure4}(c). 

Therefore, we conclude that  the dephasing between the chirality of the Dirac electrons in the top and bottom graphene layers is responsible for the existence of circular dichroism in TBG.

\section{Tight-binding results}
In the continuous model, the tunneling between layers is vertical and there is not any current in the $x$ or $y$ directions associated with the interlayer coupling. 
However, because of the twist angle, atoms in the top and bottom layers are in general not vertically stacked \footnote{To be more precise, only one pair of atoms per unit cell is  vertically aligned in a commensurate TBG.}. 
Therefore, it is intuitively appealing to think that the resulting chiral arrangement of atoms in TBG may produce 
 a contribution from the top-bottom current in the $x$ or $y$ directions. 
Such contribution would amount to a solenoid-like helical current 
encompassing 
the top and bottom graphene layers, 
 adding up 
 to the CD in TBG. 
 
To check the importance of this effect we have performed microscopic tight-binding (TB) calculations of the Hall conductivity.
We study twist angles for which the moir{\'e} pattern is commensurate with the graphene lattice parameters \cite{Lopes_2012}. 
 In the TB model we include only a $p_z$ carbon orbital per site; within each graphene sheet we only consider a nearest-neighbor hopping $\gamma_0=-2.7$ eV and 
   we assume that the interlayer coupling decreases exponentially with the separation between atoms \cite{Morell_2010,Suarez-Morell:2014aa},
\begin{equation}
H_{T,B} =\gamma_1  \sum _{i,j}  e ^{-\beta (r_{ij}-d) } c_i^+ d_j + {\rm H.c.}
\label{tunneling}
\end{equation} 
where  $d=3.35$ \AA, $\gamma_1=-0.39$ eV is the vertical interlayer hopping for AB stacking, $\beta$=3, $r_{ij}$ is the distance between atom $i$ in T layer and atom $j$ in B layer and $c_i$($d_j$) annihilates an electron in T (B) layer at site $i$ ($j$). Within this model, each atom in a layer  tunnels to the atoms in the adjacent layer located inside a circumference  with a radius several times larger than the intralayer nearest-neighbor distance between carbon atoms. Therefore, the coupling between the T and B layers includes 
tunneling between atoms 
which are not one on top of the other, i.e., they are not vertically aligned. 
When computing the current operator in the in-plane $\alpha$-direction, besides the current in the T and B layers,  there are terms 
related to non-vertical interlayer tunneling that take the form \cite{Graf:1995} 
\begin{equation}
J_{\alpha} ^{TB} =i  \frac e {\hbar} \gamma_1 \sum _{<ij>} e ^{-\beta (r_{ij}-d) }  ({\bf R} _j -{\bf R}_i)_{\alpha} c_i^+ d_j  + {\rm H.c.}\, \, 
\end{equation}
where ${\bf R} _i$ is the vector position of atom $i$. 
Note that tunneling between atoms located one on top of the other does not contribute to the current in the in-plane directions.

\begin{figure}[htbp]
\centering
\includegraphics[width=9cm,clip]{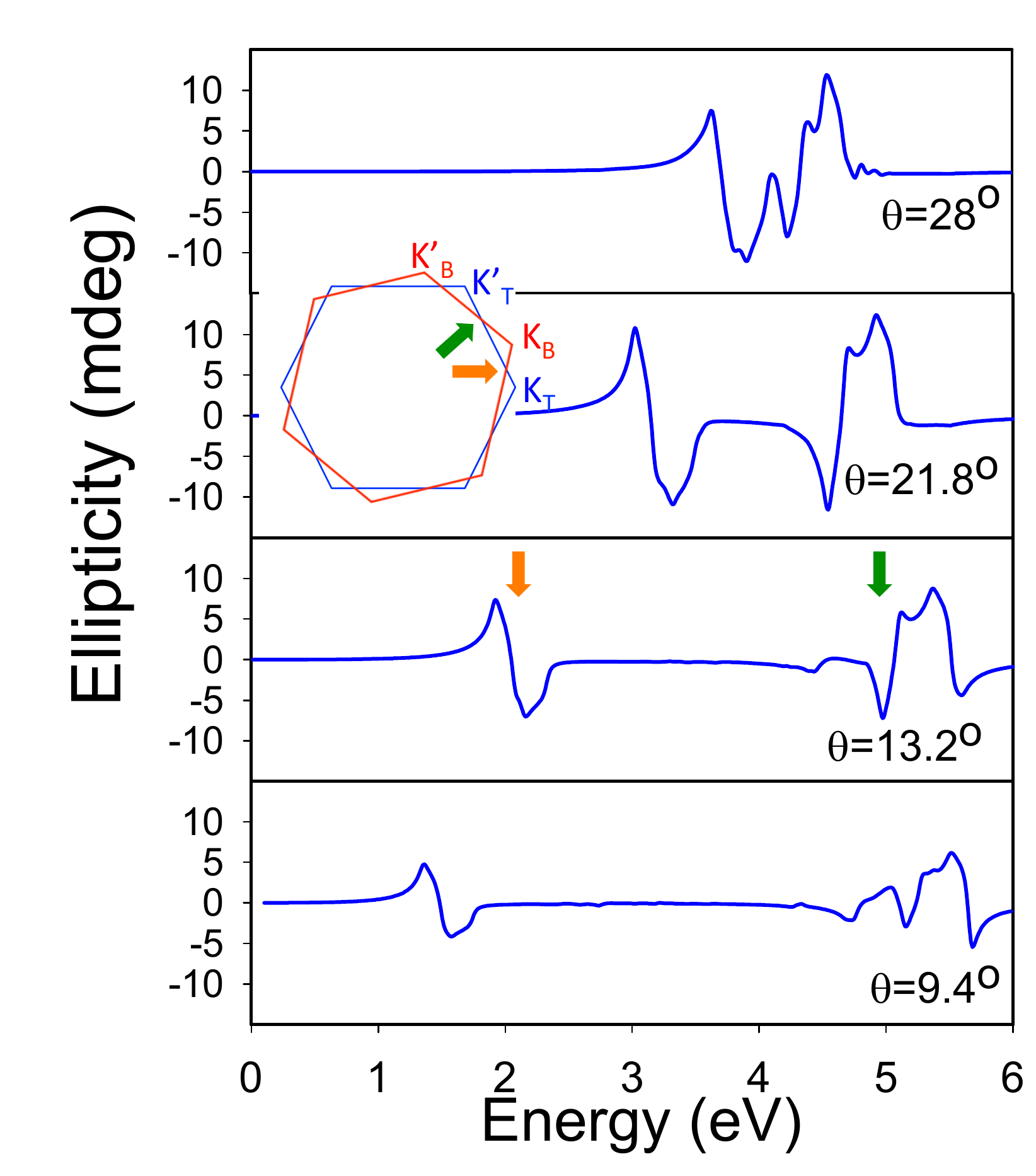}
\caption{(Color online)  Ellipticity for a TBG  for different rotation angles obtained with the tight-binding Hamiltonian. The low-energy features are related to transitions in the midpoint between equivalent rotated Dirac points $K_T$ and $K_B$ and $K'_T$ and $K_B'$. The high-energy structure 
originates from transitions near the midpoints between inequivalent rotated Dirac  points $K_T$ and $K' _B$ and $K'_T$ and $K'_B$. In the calculation we set $\eta$=0.025 eV.} 
\label{Figure5}
\end{figure} 
 
We plot in Fig. \ref{Figure5} the ellipticity obtained from tight-binding calculations for TBG with different rotation angles. The tight-binding model not only describes the linear dispersion near the Dirac points, but also the high and low-energy spectrum.
Therefore, apart from the low-energy features associated to the band anticrossing near the  equivalent $K_T$ and $K_B$ rotated Dirac points,
 another structure at higher energies appears, related to van Hove singularities at regions of the Brillouin zone corresponding to midpoints between inequivalent  Dirac points $K_T '$ and $K_B$. The low-energy structure has the same  peak-dip form than that obtained in the continuous Dirac-like Hamiltonian.  This confirms the correctness of the approximations we are using.  Note the similarity between the tight-binding and continuum results of Figs. 3 (a) and the bottom panel of Fig. 5, corresponding to $10^{\rm o}$ and $9.4^{\rm o}$, respectively.

 In order to assess the importance of the helicoidal tunneling between layers, we have calculated the ellipticity neglecting the current associated with non-vertical hopping between layers, keeping the connection between layers in the Hamiltonian. We have obtained that this contribution is always smaller than 1\%, and thus confirm that circular dichroism in TBG has its origin in the 
 rotation of the chirality 
of Dirac electrons  in the top and bottom layers.

\section{Discussion and Conclusions}
We have developed a theory for circular dichroism  in chiral bilayer structures.
In quasi-two-dimensional systems the circular dichroism is nearly proportional to the 
imaginary part of the Hall conductivity. 
Starting from the Kubo formalism for the current-current response function,
we have obtained in a natural way an expression for the imaginary part of the Hall conductivity. We find that in lowest order, it is proportional
to the phase difference of the vector potential in the top and bottom layers, $q_zd$, where $q_z$ is the wave vector of the incident light and $d$ is the distance between graphene layers.  
In a bilayer system, the Hall conductivity  is a measure  of    the  current induced  in a layer in a given direction 
as response to a perpendicular current dephased  in $q_zd$ in the opposite layer.

We  have applied this formalism to twisted bilayer graphene.  Analyzing the symmetries of
the continuous Dirac-like Hamiltonian of the bilayer, 
we identify as the source of the optical activity the rotated helicities of the coupled Dirac fermions in the top and bottom layer. Such rotation  is responsible for the peak-dip shape of the circular dichroism spectra. 
For low energies and small rotation angles ($< 10^{\rm o}$) , the accuracy of this continuum model has been verified repeatedly, reproducing crucial features such as the velocity renormalization of the carriers and the appearance of van Hove singularities for low twist angles. Furthermore, the continuum model permits semianalytical calculations that allow for a deeper understanding of the shape and structure of the optical spectra. 

We have also verified our results by means of a microscopic tight-binding calculation, that also reproduces the higher-energy features 
 of large-angle TBG, which are the pertinent cases for comparison to experiments.  
 Tight-binding calculations also  confirm the irrelevance of helical tunneling processes for the CD in twisted bilayer graphene. 
 In the low energy region, near the Dirac points, both models give similar results, so it is more advantageous to use the continuum one, especially for low angles, for which a tight-binding approach is more demanding computationally and can be unaffordable due to the large units cells. In fact, in atomistic models, only commensurate unit cells can be considered, i.e., the angle takes discrete values, while in the continuum model it can be smoothly varied.

Our approach is valid to any multilayered quasi-two-dimensional Dirac heterostructure, made of any combination of the so-called van der Waals materials, for which reflection and inversion symmetries are broken and a chiral optical response is expected. 
This could be BN on graphene or any twisted transition metal dichalcogenides such as WSe$_2$, MoS$_2$. etc. In fact, a graphene trilayer with a twist will also show this effect, so low-symmetry trilayers can also be studied along the same lines. Additionally, the excitonic spectrum in van der Waals' coupled transition metal dichalcogenides could also present circular dichroism. 
Besides being of broad applicability, our  procedure provides a unique insight on the origin of circular dichroism in twisted bilayers. 

In agreement with experiments \cite{Kim:2016aa}, our results show CD peaks at the energies corresponding to the van Hove singularities that occur in twisted bilayer graphene. The comparison between our calculations and the experiments is rather good with respect to the positions of the peaks. Furthermore, the intensity of the CD we obtain is of the order of magnitude of the experimental results, in fact, somewhat larger. The precise value could be fitted by slightly varying some parameters of the model, such as the hoppings, dielectric constant or interlayer distance. 
Nevertheless, as our interest is not to fit the experimental data, but to understand the origin of CD in TBG, we opted to take standard values for these parameters. 
However, there are 
differences in the shapes of the peaks between our theory and the experiments. In fact, the theoretically calculated spectra present more features than the experimental results. 
Additionally, the peak-dip features that appear in the theoretical CD spectra and that are relevant to explain the origin of the effect are not resolved in experiments, 
which 
only 
show 
rather broad peaks.  
Such a disparity could be due to several factors: for example, charge inhomogeneities, distinct angle domains, small potential differences across the layers, local doping, many-body or thermal effects \cite{Alivisatos:1988,Havener:2014aa}. 
It could be possible to fit the experimental curves by using an inhomogeneous Gaussian broadening, that for example in Hidalgo et al. \cite{Hidalgo:2009aa} and Kim et al.  \cite{Kim:2016aa} was as large as 0.4 eV. 
  Such broadening would mask the shape of the spectra, hiding part of the behavior that will be surely resolved in future measurements. In our case, it may subdue the peak-dip structure that we have found to be characteristic of dichroic spectra, so we prefer not to include it, having in mind that some of its sources might be eliminated in the future.  
 It should be interesting to make further experimental studies 
improving the 
resolution 
so that the CD peak structure could be verified.

\ack
ESM acknowledges financial support by Chilean FONDECYT grant 11130129. 
LC and LB  acknowledge financial support by the Spanish
MINECO Grant No. FIS2015--64654-P. We are grateful to G. G\'omez-Santos and T. Stauber for helpful discussions. 

 \section*{References}
 
 \providecommand{\newblock}{}


\bibliographystyle{iopart-num.bst}

\end{document}